\documentclass[reprint,pra,superscriptaddress,showpacs]{revtex4-1}
\usepackage{amsbsy}
\usepackage{amsfonts}
\usepackage{amssymb}
\usepackage{amsmath}
\usepackage{graphicx,color}
\usepackage{graphicx}

\begin{document}

%opening

\title{Continuous Quantum Error Correction Through Local Operations}

\author{Eduardo Mascarenhas}
\affiliation{Departamento de F\'isica, Universidade Federal de Minas Gerais, 31270-901, Belo Horizonte, Brazil}
\affiliation{Centre for Quantum Technologies, National University of Singapore, 2 Science Drive 3, 117542 Singapore}
\author{Breno Marques}
\affiliation{Departamento de F\'isica, Universidade Federal de Minas Gerais, 31270-901, Belo Horizonte, Brazil}
%
%\author{Daniel Cavalcanti}
%\affiliation{Wandering Around The Globe}

\author{Marcelo Terra Cunha}
\affiliation{Departamento de Matem\'atica, Universidade Federal de Minas Gerais, 30123-970, Belo Horizonte, Brazil}
\author{Marcelo Fran\c{c}a Santos}
\affiliation{Departamento de F\'isica, Universidade Federal de Minas Gerais, 31270-901, Belo Horizonte, Brazil}
\affiliation{Centre for Quantum Technologies, National University of Singapore, 2 Science Drive 3, 117542 Singapore}

\begin{abstract}
We propose local strategies to protect global quantum information. The protocols, which are quantum error correcting codes for dissipative systems, are based on environment measurements, direct feedback control and simple encoding of the logical qubits into physical qutrits whose decaying transitions are indistinguishable and equally probable. The simple addition of one extra level in the description of the subsystems allows for local actions to fully and deterministically protect global resources, such as entanglement. We present codes for both quantum jump and quantum state diffusion measurement strategies and test them against several sources of inefficiency. The use of qutrits in information protocols suggests further characterization of qutrit-qutrit disentanglement dynamics, which we also give together with simple local environment measurement schemes able to prevent distillability sudden death and even enhance entanglement in situations in which our feedback error correction is not possible.
\end{abstract}
\pacs{ 03.65.Yz; 03.67.Hk; 03.67.Pp}
\maketitle

\section{Introduction}

Every physical system is embedded in a neighborhood of other systems, also known as its environment.
Generically, the system exchanges energy and information with the environment through quantum interaction processes which lead to system-environment correlations. If we cannot keep track of the environmental degrees of freedom we eventually lose information about the system on so-called decoherence processes~\cite{Deco}. This is a major obstacle for quantum computation and quantum information processing, since this environmental interaction spoils quantum data, in the sense that it disrupts qubit states causing errors and degrades resources such as entanglement, necessary for some quantum algorithms. In the quantum scale the environment was long seen as an entity whose action needed to be eliminated in order to preserve quantum information in the system. 

As an attempt to address this issue, quantum error correcting codes were developed to suppress the degradation induced by the environment~\cite{Shor,Steane,Laflamme,Bennett,Gottesman}. Most of these codes rely on redundancy and global operations, meaning they use many physical qubits to encode the logical qubits and use global operations on these several qubits to correct for the errors and/or detect the error syndromes. Passive codes, named error avoiding codes, were also developed and those are based on decoherence free subspaces in which a subspace of the system does not evolve under the coupling to the environment~\cite{Eavoid}. The first designed error correcting codes used projective measurements to detect the errors and unitary operations to correct them. Not long after the first codes, and still recently, environment measurements and Hamiltonian feedback were employed in error correction protocols~\cite{Mabuchi,Milburn,Plenio,JumpECorr,CECF,CECW,PEC,EFC}. In the context of errors caused by spontaneous emission, hybrid codes combining passive and active codes were developed \cite{Mabuchi,Milburn,Plenio,JumpECorr}, in which there are clearly two different kinds of errors, those that are avoided by encoding the quantum data in a decoherence free subspace and those that need a direct action given that the error has occurred. Such spontaneous emission correcting codes relied on the ability to continuously monitor the environment (global measurements on independent environments in~\cite{Mabuchi} or a measurement over a global environment in~\cite{Plenio}), on the encoding of the logical qubits into entangled physical qubit states and also on complicated global operations on several subsystems for the active error correction. It was then realized that a significant redundancy reduction was possible when the distance between the physical systems is much larger than the wavelength of the emitted excitations~\cite{JumpECorr}, which also makes the error detection a local process. Further improvement on the number of required resources was given in~\cite{CECW} with the aid of a global driving Hamiltonian. Therefore statistically independent environments probably provide a better suited scenario for less demanding quantum computation. However, these proposals scale badly in the sense that as the number of logical qubits grows one needs to collectively control the behavior of many systems through intricate global operations in order to protect quantum information which is encoded into multi-qubit entangled states. As another possible strategy, continuous feedback has been applied to suppress decoherence due to spontaneous emission~\cite{FeedDeco}, and feedback based strategies have also been devised to create and protect known entangled states~\cite{FeedEnt}. However, all the previous efficient proposals for sustaining entanglement require global operations and hence do not allow for distant entanglement based communication~\cite{FeedEnt}.

Let us clarify what we mean by local and global operations. Consider a network whose nodes are quantum systems, and that each quantum system may be composed of smaller subsystems: in our case three level systems (or qutrits). By global operation we mean any operation that is not a product of single operations on the individual subsystems, and conversely, a local operation can be expressed as a product of such single location operations. Local operations such as single qudit operations in unison can be simple to implement, however operations such as measurements comprising more than 2 close-by qudit locations can be very hard to implement. Furthermore, if the subsystems are located at distant nodes global operation might be impossible to implement. Therefore, error correcting codes relying solely on local operations are much more desirable.

In~\cite{Nos} the authors have proposed a \textit{local decay quantum error correcting code} also combining active and passive protocols in which a logical qubit is locally encoded in a physical qutrit. In this code the errors are detected by locally measuring the independent environments and the system is deterministically recycled through local feedback. Therefore the code is $100\%$ local and requires neither entangled ancillas nor global operations. Similar qudit encoding techniques were employed for error correcting continuous variable systems and phase dumping in qudit systems~\cite{EqubitOsc,QditCode}. However, for the spontaneous emission codes shown here and in~\cite{Nos}, the minimal qudit dimension is $d=3$, which is much less demanding and much simpler than the codes shown in~\cite{EqubitOsc,QditCode}. Since the protocol is strictly local it can also be used to protect known entangled states shared by spatially separated different parties, thus allowing for entanglement based quantum communication between them.

In this paper we extend these results to the diffusion limit, that is, we develop a similar qutrit encoding and feedback error correction for the case in which the measurements performed  over the reservoirs present extra sources of noise other than the system-environment interaction (such as that introduced by a classical oscillator in a homodyne measurement).
We also extend the analysis previously done, testing our protocol against several sources of inefficiency and pointing out the most important ones. 

Finally, it is worth saying that this feedback error correction is based on the indistinguishability of the decay channels of the qutrits in a cascade-like energy structure. In the end of the paper, we analyze the dynamics of qutrit-qutrit entanglement for different energy configurations of the three-level systems and analyze the possibility of maintaing or even enhancing entanglement through reservoir monitoring, also extending earlier results on the effects of abrupt changes and sudden death of distillability~\cite{Abrupt} in $3\otimes 3$ systems.

\section{Measuring The Environment and Information Feedback}

We consider a global system composed of internal subsystems that are weakly and
smoothly coupled to their own local and independent reservoirs.
These couplings should respect typical Markov and Born approximations~\cite{Heinz, Carmichael}. We assume that the system is prepared in a given state (usually a pure state) and that it is not correlated to the environment at this stage. This can be expressed as $|\Psi(0)\rangle=|S\rangle|E\rangle$, with $S$ and $E$ (which will be taken as the vacuum) designating system and environment initial states, respectively.
Correlations arise as a result of the evolution process that generically can be given by $|\Psi(t)\rangle=\sum_i \sqrt{p_i} |S_i(t)\rangle|E_i(t)\rangle$. However, in deriving the theoretical model one assumes that the environment is almost unaffected by the interaction (as a thermal bath, in our case at zero temperature) and that the state $\Psi(t)$ presents correlations only to first order in the interaction coupling. Another important point is that the environment quickly loses memory of its past history as compared to the time it takes for the system to evolve appreciably. The interaction Hamiltonian can be written in the interaction picture under the rotating wave approximation as
\begin{equation}H_{int}=\sum_ki\left[\Pi_k b_k^{\dagger}(t)-\Pi_k^{\dagger}b_k(t)\right],\end{equation}
in which statistically independent reservoirs are represented by $b_k(t)=\frac{1}{\sqrt{2\pi}}\int_{-\infty}^{\infty}d\omega b_k(\omega)e^{-i(\omega-\omega_{k})t}$, 
with $b(\omega)$ the environment many degrees of freedom labeled by frequencies $\omega$ and $\Pi_k$ are system operators with characteristic frequencies $\omega_k$. The coupling strength is already included in the definition of the system operators. The global state vector evolution can be given by the quantum stochastic Schr\"{o}dinger equation~\cite{Gardiner}
\begin{equation}
(\textbf{S}) \quad d|\Psi(t)\rangle=\sum_k\left\{\Pi_k dB_k^{\dagger}(t)-\Pi_k^{\dagger}dB_k(t)\right\}|\Psi(t)\rangle,
\end{equation}
with the quantum white noise or quantum Wiener processes $B_k(t)=\int_0^tb_k(s)ds$. Therefore with independent Markovian reservoirs initially at the vacuum state we have the operator averages
$\langle dB_{k}(t)dB_l^{\dagger}(t')\rangle=\delta_{kl}\delta(t-t')dt$, and $\langle dB_{k}^{\dagger}(t)dB_l(t')\rangle=\langle dB_{k}^{\dagger}(t)\rangle=\langle dB_{k}(t)\rangle=0$. This equation is in the Stratonovich form and it is convenient to convert it to the Ito form with the corresponding Ito rules $dB_{k}(t)dB_l^{\dagger}(t')=\delta_{kl}\delta(t-t')dt$, and $dB_{k}^{\dagger}(t)dB_l(t')=[dB_{k}^{\dagger}(t)]^2=[dB_{k}(t)]^{2}=0$, yielding
\begin{equation}
(\textbf{I})\quad d|\Psi\rangle=\sum_k\left\{-\frac{1}{2}\Pi_k^{\dagger}\Pi_kdt +\Pi_k dB_k^{\dagger}-\Pi_k^{\dagger}dB_k\right\}|\Psi\rangle.\label{QSSEI}
\end{equation}
Now, if we are ignorant with respect to the environment and focus solely on the system properties our knowledge of the system is described by the reduced state $\overline{\rho(t)}=\mathrm{tr}_{E}\{|\Psi(t)\rangle\langle\Psi(t)|\}$, whose dynamics is given by the master equation
\begin{equation}
d\overline{\rho}=\mathcal{L}[\Pi]\overline{\rho}dt=\sum_k\left\{-\frac{1}{2}\{\Pi_k^{\dagger}\Pi_k,\overline{\rho}\}+\Pi_k\overline{\rho}\Pi_k^{\dagger}\right\}dt,\label{ME}
\end{equation}
in which the over-line denotes an average over the reservoir. By ignoring the environment, either willingly or not, we also lose information on the system and this evolution leads to loss of purity and entanglement of the initially prepared state.

\subsection{Measuring The Environment}

We may choose to perform continuous selective measurements on the environment instead of averaging over it. By \textit{selectively collecting} the measurement results we acquire useful information. Take the initial state $|\Psi(0)\rangle=|S\rangle|\textbf{0}\rangle$, with the reservoir in the vacuum state (only one reservoir for simplicity) and assuming there is nonzero excitations in the system. If the interaction Hamiltonian $H_{int}$ allows for excitation exchange, after a very small time the global system evolves to $|\Psi(dt)\rangle=\sqrt{p_0}|S_0\rangle|\textbf{0}\rangle+\sqrt{p_1}|S_1\rangle|\textbf{1}\rangle$, with $p_{1(0)}$ the probability of one (no) excitation reaching the environment. This can be seen by directly iterating equation~(\ref{QSSEI}), with $|S_0\rangle=[\openone-\frac{1}{2}\Pi^{\dagger}\Pi dt]|S\rangle/\sqrt{p_0}$, and $|S_1\rangle=\Pi \sqrt{dt}|S\rangle/\sqrt{p_1}$, with $p_1=dt\langle S|\Pi^{\dagger}\Pi|S\rangle$. Equation~(\ref{ME}) gives us the system state $\overline{\rho}(dt)=p_0|S_0\rangle\langle S_0|+p_1|S_1\rangle\langle S_1|$, while measuring the environment gives us $|S_{0(1)}\rangle$ with probability $p_{0(1)}$. This corresponds to measuring the number operator in the environment $d\Lambda(t)$, with $\Lambda(t)=\int_0^tb^{\dagger}(s)b(s)ds$, whose eigenvalues are the number of excitations counted in a time interval $dt$, one or zero excitations being the only possibilities~\cite{Gardiner,Bart}. Such measurement scheme is called a quantum jump unraveling~\cite{Jump}. In principle, there are infinitely many other ways of observing the environment. In particular, one of them is the quadrature operator $d\Theta=(\Pi^{\dagger}+\Pi)dt+dB^{\dagger}(t)+dB(t)$ or homodyne measurement that gives rise to a diffusive unraveling~\cite{Diffusion}. Diffusive unravelings are suitable in optics when one combines a system output with a classical local oscillator on a beam splitter and measures the resulting output. They arise naturally in condensed matter physics, for instance in quantum dot monitoring with a quantum point contact or single electron transistor~\cite{Solid}, and are also very useful for entanglement~\cite{Andre} and geometric phase~\cite{GeometricSerbia} characterization in open quantum systems.

The quantum measurement results are, ultimately, classical records and therefore can be expressed by classical (instead of quantum) stochastic processes. The eigenvalues for the measurement operators are a Poisson process $dN(t)$ for the quantum jump (where $dN(t)$ assumes values $0$ or $1$) or, in the case of the diffusive unravellings, $dQ(t)=\langle \Pi^{\dagger}+\Pi\rangle dt+dW(t)$, where $dW(t)$ is a Wiener process~\cite{Graham}. Such processes obey the usual Ito rules $dN_k(t)dN_l(t')=\delta_{kl}\delta(t-t')dN_k(t)$, $dW_k(t)dW_l(t')=\delta_{kl}\delta(t-t')dt$~\cite{Stochastic}. 
Thus, the conditional evolution of the system (where tilde indicates unnormalized state) can be written as a function of the measurement records 
\begin{equation}d\widetilde{\rho}(t)=\sum_k\bigg\{\mathcal{A}_{\xi_k}[\Pi_k]dt+\mathcal{B}_{\xi_k}[\Pi_k]d\xi_k(t)\bigg\}\widetilde{\rho}(t),\label{Conditioned}\end{equation}
with $\xi(t)$ a classical stochastic function that records the measurement results, and represents the information we acquire from observing the environment. $\mathcal{A}_{\xi}[\Pi]$ and $\mathcal{B}_{\xi}[\Pi]$ are superoperators acting on the density matrix of the system. The operators $\Pi$ can now be interpreted as a set of detectors surrounding the system in the environment. For a jump unraveling we have
$\mathcal{A}_{N_k}[\Pi_k]\rho=-\frac{1}{2}\{\Pi_k^{\dagger}\Pi_k,\rho\}$, and
$\mathcal{B}_{N_k}[\Pi_k]\rho=\left[\Pi_k\rho\Pi_k^{\dagger}-\openone\rho\right]$, with the Poisson process $d\xi_k=dN_k$ assuming the value one if the $k$th detector is triggered and zero if not, with mean value $\overline{dN}=\langle \Pi^{\dagger}\Pi\rangle dt$. For a diffusive unraveling
$\mathcal{A}_{Q_k}[\Pi_k]=\mathcal{L}[\Pi_k]$, and
$\mathcal{B}_{Q_k}[\Pi_k]\rho=\rho\Pi_k^{\dagger}+\Pi_k\rho$, with the $k$th detector recording the intensity $d\xi_k=dQ_k$ and the Wiener process assuming any real value with a zero mean Gaussian distribution of width $dt$. Every detection apparatus suffers from imperfections, which leads to a non unit efficiency $\eta$ not accounted for in the above operators. Including inefficiencies changes the jump unravelling into $\mathcal{A}_{N\eta_k}[\Pi_k]\rho=\mathcal{L}[\Pi_k]\rho-\eta_k\Pi_k\rho\Pi^{\dagger}_k$, with a consequent decrease in the probability of triggering the detectors reflected in the mean value $\overline{dN}=\eta\langle \Pi^{\dagger}\Pi\rangle dt$. Whereas, the diffusive unraveling changes as $dQ_k=\eta_k\langle\Pi_k^{\dagger}+\Pi_k\rangle dt+\sqrt{\eta_k}dW$. This approach to inefficient detection is based only on an information perspective, in the sense that it describes an ability of distinguishing the results that decreases with the efficiency. Realistic detection models can be found for specific systems in~\cite{Realistic}.

\subsection{Information Feedback}

Now suppose that in a practical situation we wish for the system to behave in a controlled way, for instance that its state, either known or not, to be preserved. One strategy for obtaining such control over the system is quantum feedback in which the observer acts on the system accordingly to the measurement outcomes~\cite{MFeedback,BFeedback}. Given the outcomes $d\xi(t-\tau)$ at the time $t-\tau$ the feedback apparatus acts with a Hamiltonian proportional to an operator $F$ at time $t$. The apparatus takes the delay time $\tau$ to process the measurement result and activate the feedback Hamiltonian. One straightforward way~\cite{MFeedback} of doing this is with a Hamiltonian linear in the measurement outcomes $H_{\textrm{fb}}=F\frac{d\xi(t-\tau)}{dt}$. With the feedback acting after the measurement we have the evolution of the system given by (to be interpreted in the Ito calculus)
\begin{equation}d\rho(t)=\Bigg\{e^{\mathcal{F}d\xi(t-\tau)}\bigg[\openone+\mathcal{A}_{\xi}[\Pi]dt+\mathcal{B}_{\xi}[\Pi]d\xi(t)\bigg]-\openone\Bigg\}\rho(t),\end{equation}
with $\mathcal{F}\rho=-i[F,\rho]$.
\begin{figure}[h]
\includegraphics[width=6cm]{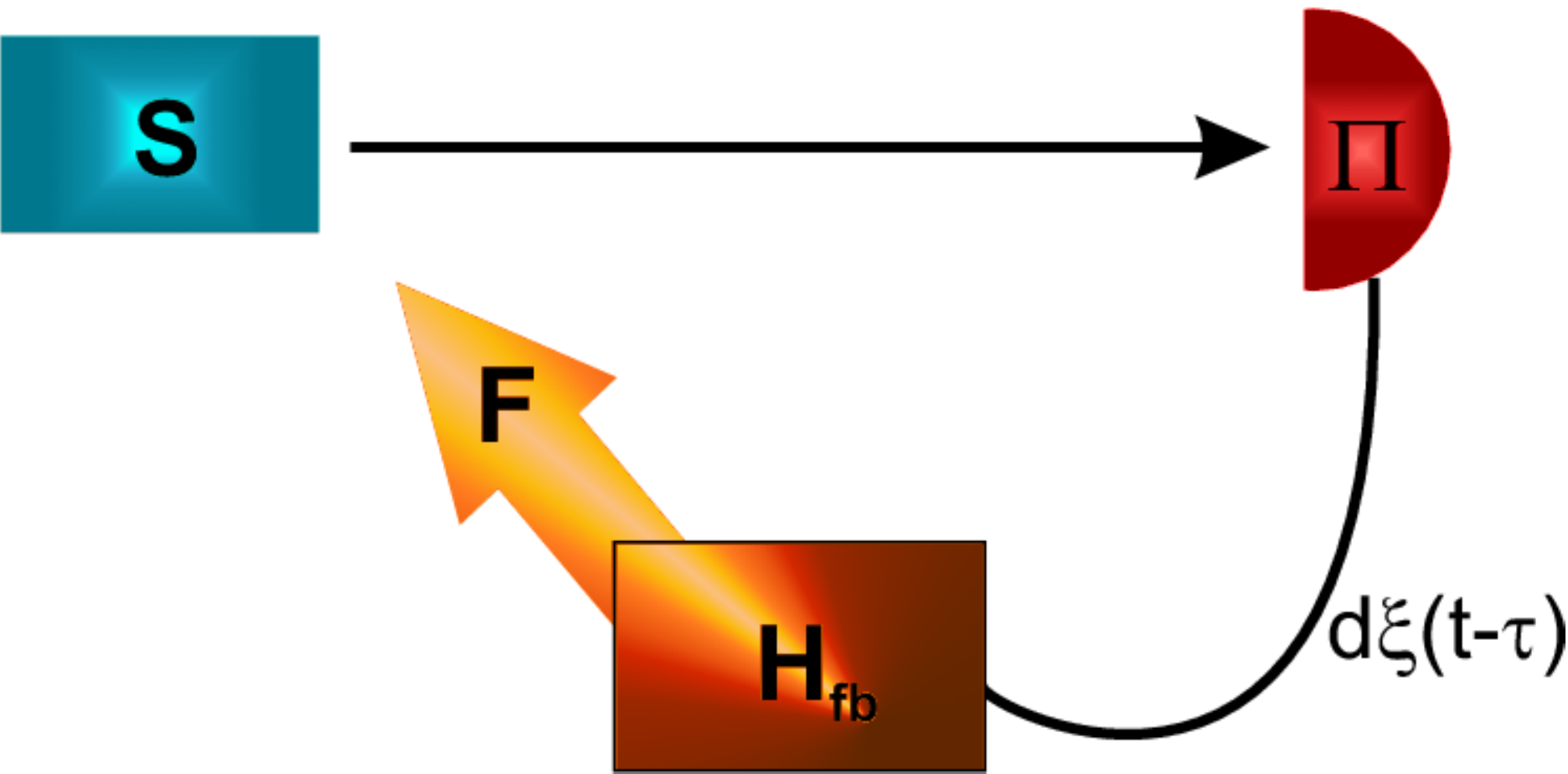}
\caption{Representation of the feedback loop. The system emits a signal which is recorded in the photocurrent $I(t)=\frac{d\xi(t)}{dt}$ by the detector $\Pi$. The measurement record than dictates the feedback action by the Hamiltonian $F$.}
\end{figure}

For example, in the quantum jump scenario, the feedback acts unitarily and in a point like manner. For non unit measurement efficiency $\eta$ and nonzero feedback time delay the dynamics will then be given by
\begin{equation}d\widetilde{\rho}(t)=\Bigg\{\bigg[e^{\mathcal{F}}-\openone\bigg]dN(t-\tau)+\mathcal{L}dt
+\mathcal{B}_{N}dN(t)\Bigg\}\widetilde{\rho}(t)-\eta\Pi\widetilde{\rho}(t)\Pi^{\dagger} dt.\label{JumpFeed}\end{equation}
Whereas, in the diffusive unraveling the dynamics will be corrected by inefficiencies to 
\begin{equation}d\widetilde{\rho}(t)=\Bigg\{\frac{\mathcal{F}}{\eta}dQ(t-\tau)+\left[\frac{\mathcal{F}^{2}}{2\eta}+\mathcal{L}\right]dt+\mathcal{B}_{Q}dQ(t)\Bigg\}\widetilde{\rho}(t).\end{equation}

\section{Local Recycling}

\subsection{Quantum Jumps}

Let us start by reviewing some basic concepts of standard quantum error correction~\cite{Laflamme}.
First we choose a subspace, a codespace, of the system to encode the logical qubits. The system is subjected to an environment induced error $\mathcal{E}$ that can be represented in a short time expansion by $\mathcal{E}\rho=[\openone+\mathcal{L}dt]\rho=\sum_{i}\Gamma_i\rho\Gamma^{\dagger}_i$. The system can be recycled after errors have occurred if there exists a recovery operation $\mathcal{R}$ such that $\mathcal{RE}\rho=\rho$
for all states in the codespace $\rho=P_C\rho P_C$, with $P_C$ the projector onto the codespace. We are interested in errors induced by a Markovian environment, and for the simplest case we have the errors $\Gamma_0=\openone-\frac{1}{2}\Pi^{\dagger}\Pi dt$ and $\Gamma_1=\Pi\sqrt{dt}$. Standard error correction should then be applied as discrete actions over time intervals to be made very small and it has to simultaneously correct for different errors $\Gamma_i$. By measuring and counting the excitations in the environment we can try to design specific recycling operations for each of the possible errors, moreover we can even choose a codespace which is decoherence free with respect to at least one error so we do not have to correct for all of them. We find a codespace which is an eigenspace of the $\Gamma_0$ operator and we have to recycle the system only when a quantum jump occurs. We need to find an operation $\mathcal{R}$ that reverses the quantum measurement and we choose to do it deterministically instead of probabilistically~\cite{ProbReverse}.

Let us then recall the conditions for deterministic reversibility of a quantum jump~\cite{Mabuchi,NielsenCaves}. The recycling operation is an unitary $R$ such that $R\Pi\propto P_C$ (without loosing generality we neglect any effect of both the recycling and jump operation outside the codespace), therefore $\Pi \propto R^{\dagger}P_C$. Note that this implies that the codespace has to be an eigenspace of the $\Gamma_0$ operator since  $\Pi^{\dagger}\Pi\propto P_CRR^{\dagger}P_C=P_C$. Therefore, in a quantum jump unraveling, one can deterministically restore the quantum data in the codspace with a direct feedback mechanism with Hamiltonian $H_{\mathrm{fb}}=F\frac{dN}{dt}$ and recycling operation $R=e^{-iF}$, if the detection operator $\Pi$ obeys the above conditions. The feedback term in equation~(\ref{JumpFeed}) can be identified with the reversing operation by $e^{\mathcal{F}}\rho=\mathcal{R}\rho=R\rho R^{\dagger}$.

\begin{figure}[h]
\includegraphics[width=8cm]{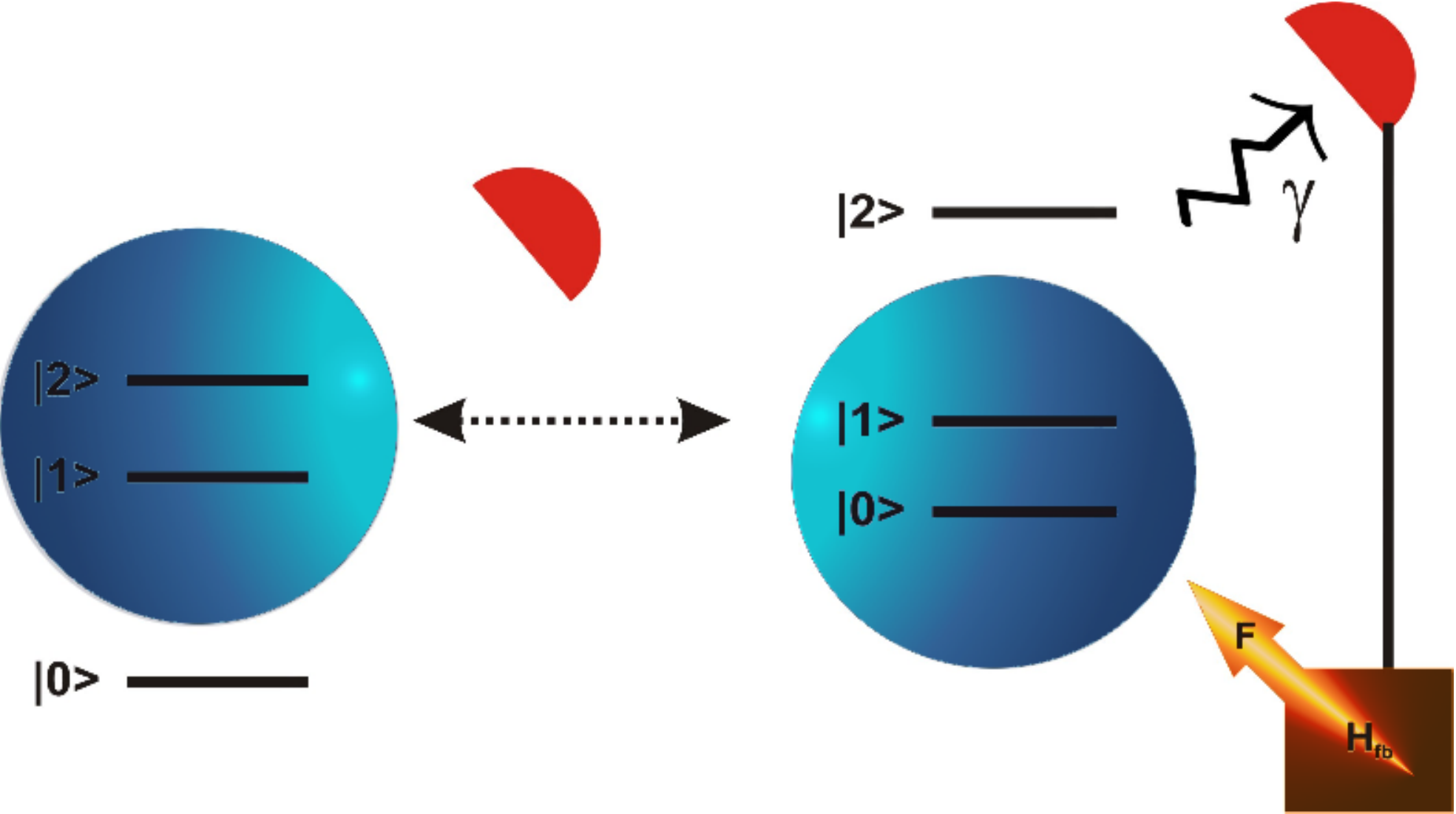}
\caption{Representation of the local recycling process. The logical qubits state is represented by the blue circle around the qutrit states. The state is preserved while the system does not emit an excitation. When in does the encoded state shifts down the qutrit states and is recycled back to its original subspace by the feedback action.}\label{Recycle}
\end{figure}

Now we turn to the specific example of locally protecting quantum information against spontaneous emission. 
Suppose we have physical qutrits each with levels $\{|0\rangle,|1\rangle,|2\rangle \}$ with  jump operators $\Pi=\sqrt{\gamma}(|1\rangle\langle2|+|0\rangle\langle1|)$. In such decay process, by detecting the emitted excitations, we cannot tell which of the qutrit's levels decayed. This is a very important condition for the protocol to work. Another favorable condition is that the decay rates of transitions $|2\rangle\mapsto|1\rangle$ and $|1\rangle\mapsto|0\rangle$ be the same $\gamma$. In the long run all states are then dragged to the ground state. Is there a subspace in which the quantum jump can be reversed? The answer is affirmative with the codespace being $\{|1\rangle,|2\rangle\}$. To corroborate this one just needs to compute $\Pi^{\dagger}\Pi=\gamma P_C$, with $P_C=|1\rangle\langle1|+|2\rangle\langle2|$. The recycling operation is then given by $R=\frac{1}{\sqrt{\gamma}}\Pi^{\dagger}+|0\rangle\langle2|$ and the feedback operator $F=\lambda[\frac{1}{\sqrt{\gamma}}(\Pi^{\dagger}-\Pi)+|0\rangle\langle2|-|2\rangle\langle0| ]$ with $\lambda\approx1,2092i$. This reversing operation clearly feeds the system back with one excitation after it has decayed.
Information is then preserved in the codespace while the detector is not triggered.
As the system evolves it eventually decays and the qutrit state might partially or completely leave the codespace. Then the detector is triggered and the feedback mechanism activated, and the system is recycled back to the original state. If the measurement is completely efficient and the feedback time delay is null then the logical qubit is locally and perfectly protected from its coupling to the environment. The recycling process is illustrated in fig. (\ref{Recycle}). Now suppose we had $n$ qutrits, each with its own recycling process, then we could protect any state in the codespace $\{|1\rangle,|2\rangle\}^{\otimes n}$, thus protecting $n$ logical qubits encoded into $n$ individual qutrits.

\subsection{Quantum State Diffusion}

We can start analyzing imperfections in the recycling protocol by looking at a very important sort of inefficiency that manifests itself as additional noise in the measurement records, either intentionally added or not by the observer, with the measurement process corresponding to a diffusive unraveling. We refer to the diffusive limit as an inefficient limit since the additional classical noise erases part of the information regarding the underlying interaction process between system and environment, which is an excitation exchange. In the jump limit we are clearly informed whether there has been an excitation exchange or not, and this is not the case in the diffusive limit. However we may also state that we are changing the way we detect the error syndromes by changing the way we measure and acquire information from the environment, and then we show that to reach error correction the encoding and the feedback Hamiltonian are also modified. Therefore the diffusion based recycling is different to the jump based and represents a different error correction code on its own. 

Some goals were shown to be possibly achieved with quantum state diffusion such as stabilization of an arbitrary (but known) pure qubit state~\cite{qubitDiff} and even global error correction~\cite{CECW} (relying on global encoding and global recovery operations). We show that it is impossible to perfectly protect an arbitrary unknown qubit state encoded in a physical qubit.  However, by introducing the qutrit encoding we can achieve error correction for a logical qubit encoded in the qutrit. Furthermore, we globally protect an arbitrary unknown multi-qubit state
from its interaction with the environment through local diffusion based feedback. 

In the diffusive limit the unconditioned feedback master equation reads (with perfect measurement and feedback)
\begin{equation}d\rho=-\frac{i}{2}\left[\Pi^{\dagger}F+F\Pi,\rho\right]dt+\mathcal{L}[\Pi-iF]\rho dt,\end{equation}
and to protect an arbitrary state one needs to add a constant driving Hamiltonian $H=-\frac{1}{2}(\Pi^{\dagger}F+F\Pi)$ to cancel the first term and to find an appropriate feedback Hamiltonian $F$ that cancels the source of decoherence in the second term. The jump operator can be written as $\Pi=M+X+iY$, with $M$ a multiple of the identity and $X$ and $Y$ are Hermitian operators. In the particular case of spontaneous emission we have $M=0$ and $X$ and $Y$ as the renormalized  angular momentum operators for the physical qutrit $X=\frac{\sqrt{\gamma}}{2}\left( |1\rangle\langle2|+|0\rangle\langle1|+h.c. \right)$ and $Y=\frac{\sqrt{\gamma}}{2}\left( -i|1\rangle\langle2|-i|0\rangle\langle1|+h.c. \right)$.
Now suppose that the codespace has some stabilizer operator $S$, which means that the codespace is an eigenspace of $S$ with eigenvalue $+1$. Then the appropriate feedback would be given by $F=Y-iXS$ such that $\mathcal{L}[\Pi-iF]=\mathcal{L}[M+X(\openone-S)]$, and thus the codespace is annihilated by $(\openone-S)$ and is preserved throughout the evolution. However, the feedback Hamiltonian must be Hermitian and hence $S$ must anti-commute with $X$, $SX+XS=0$~\cite{CECW}. In the qubit case the only local stabilizer would be the identity operator which does not anti-commute with any other operator and thus it is impossible to find the feedback Hamiltonian and perfectly protect one qubit with diffusion based feedback.

\begin{figure}[h]
\includegraphics[width=8cm]{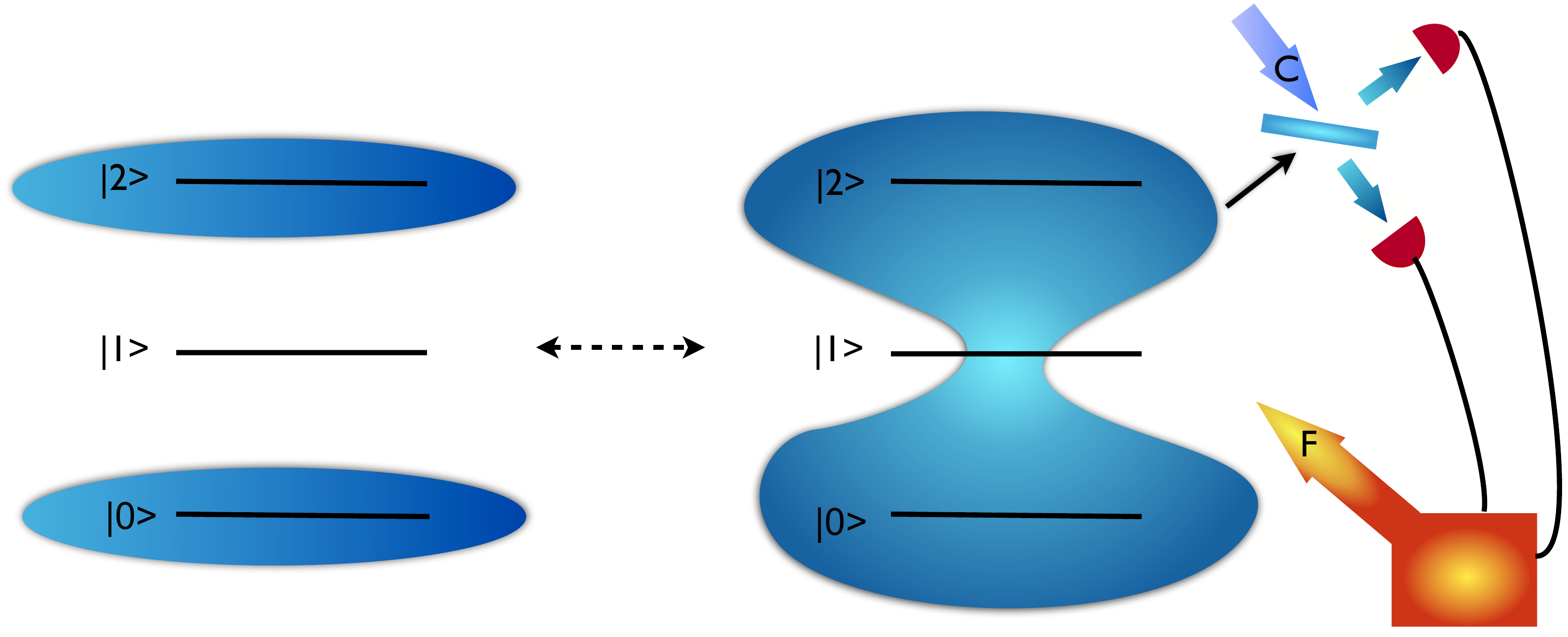}
\caption{Representation of the local recycling process. The logical qubit state is encoded in the blue region around the qutrit states. The system suffers infinitesimal errors continuously and the state is damaged in an irregular way, such that part of the information leaves the codespace and the remaining part is also disrupted.  
Then the error is detected (for example, by homodyne measurements in optical systems) and the feedback mechanism is activated and the state is recycled back to its original subspace by the feedback action.}\label{RecycleHo}
\end{figure}
In the qutrit case it is possible to find a useful stabilizer that anti-commutes with the $X$ angular momentum operator. The stabilizer is then defined by $S|0\rangle=|0\rangle$, $S|2\rangle=|2\rangle$, and $S|1\rangle=-|1\rangle$, and thus the codespace is $\{|0\rangle,|2\rangle\}$ with $P_C=|0\rangle\langle0|+|2\rangle\langle2|$. 
The recycling process goes as follows (see also figure \ref{RecycleHo}). In a very short time interval the detectors register the measurement outcome $dQ(t)$ and the system state evolves accordingly as $|S_Q(t+dt)\rangle=\left[\openone-\frac{1}{2}\Pi^{\dagger}\Pi dt+\Pi dQ(t)\right]|S_Q(t)\rangle$, which damages the initial state. However, immediately after the detection the feedback mechanism is triggered and the above state is multiplied by the feedback $R=e^{-iFdQ(t)}$, thus the system actual evolution is 
\begin{eqnarray}|S_Q(t+dt)\rangle&=&\bigg[\openone-\bigg(\frac{1}{2}\Pi^{\dagger}\Pi-i\frac{1}{2}(\Pi^{\dagger}F+F\Pi)+iF\Pi \nonumber \\
&+&\frac{F^2}{2}\bigg) dt +\left(\Pi -iF\right)dQ(t)\bigg]|S_Q(t)\rangle \nonumber \\
&=&\bigg[\openone -\left(\frac{\gamma dt}{2}+XdQ\right)(\openone-S) \bigg]|S_Q(t)\rangle.\nonumber\end{eqnarray}
Now, as we defined the appropriate feedback hamiltonian and stabilizer, it is easy to see that if the initial state is inside the codespace ($P_C|S_Q(t)\rangle=|S_Q(t)\rangle$) then it remains in the codespace $|S_Q(t+dt)\rangle=|S_Q(t)\rangle$, since the term with the stabilizer $S$ annihilates the codespace $(\openone-S)|S_Q\rangle=2(\openone-P_C)|S_Q\rangle=0$.

Therefore, as in the case of quantum jump based feedback, the encoding into qutrits allows for local actions to fully protect $n$-qubit states, but now with a diffusive environment measurement and the codespace $\{|0\rangle,|2\rangle\}^{\otimes n}$. However the diffusion based recycling can be more sensitive to inefficiencies as compared to the jump based one, at least in single realizations. When we consider inefficiencies, for instance in the feedback time delay, the trajectories become highly irregular. This behavior is a manifestation of the lack of information in the diffusive unraveling which induces degradation of entanglement. In fact diffusive unravelings 
are also know to minimize the average entanglement when we average over trajectories~\cite{Andre}.

\subsection{Inefficiencies}

Now we analyze the performance of the protocol subjected to inefficiencies. Since the error detection relies on measuring the environment the protocol heavily depends on the measurement efficiency. The feedback time delay must be significantly smaller than the decay time otherwise the systems dissipates before the feedback action. The measurement time resolution $dt$ must be much smaller than the decay time although it has to be grater than the reservoir characteristic correlation time. We focus the analysis on the jump based code since it presents interesting effects  on single realizations. 

\begin{figure}[h]
\includegraphics[width=9cm]{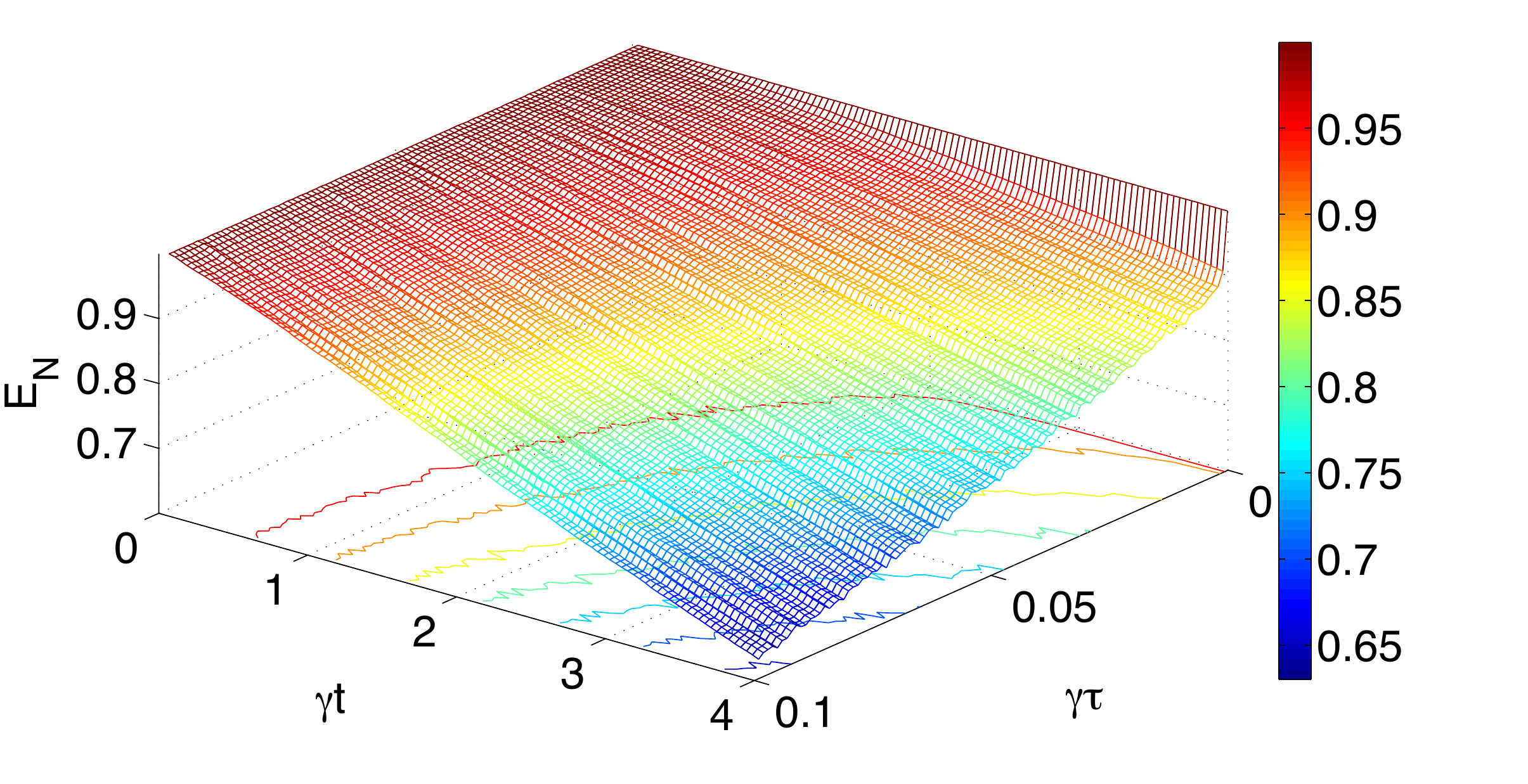}
\caption{Average entanglement (negativity) of the jump feedback as a function of time $t$ and feedback time delay $\tau$.}\label{FeedbackTimeDelay}
\end{figure}

It is clear that global resources such as entanglement can also be locally preserved by the recycling protocol. Therefore, distant parties on a quantum network can sustain entangled states against the environmental action and perform quantum communication protocols such as teleportation~\cite{Teleport} and dense coding~\cite{Dense}. For such purpose using robust states would increase the communication efficiency. For instance, states that do not suffer from entanglement sudden death. W type states are know to be highly robust and in our recycling protocol they would take the forms of $|12\rangle+|21\rangle$ and $|112\rangle+|121\rangle+|211\rangle$ for two and three qutrits respectively. There is an interesting mechanism that makes such states robust against finite feedback time delay. Let us examine the simplest case with an initial state $|12\rangle+|21\rangle$ and a feedback time delay $\tau$. Suppose that an excitation is emitted and detected from the first qutrit and the state jumps to $|02\rangle+|11\rangle$ partially leaving the codespace. Now, this state evolves as $|02\rangle+e^{-\gamma\tau/2}|11\rangle$ while the feedback action is delayed, and when it does act the state is put back into the codespace $|12\rangle+e^{-\gamma\tau/2}|21\rangle$ although not completely restored. Now, if an excitation is detected from the second qutrit the state evolves and by the end of the delay time it is again a maximally entangled state $e^{-\gamma\tau/2}[|11\rangle+|20\rangle]$, and then it is finally restored by the feedback. Thus the undesirable evolution that takes place during the feedback delay on one of the qutrits can be reversed by the delay on another qutrit.
\begin{figure}[h]
\includegraphics[width=9cm]{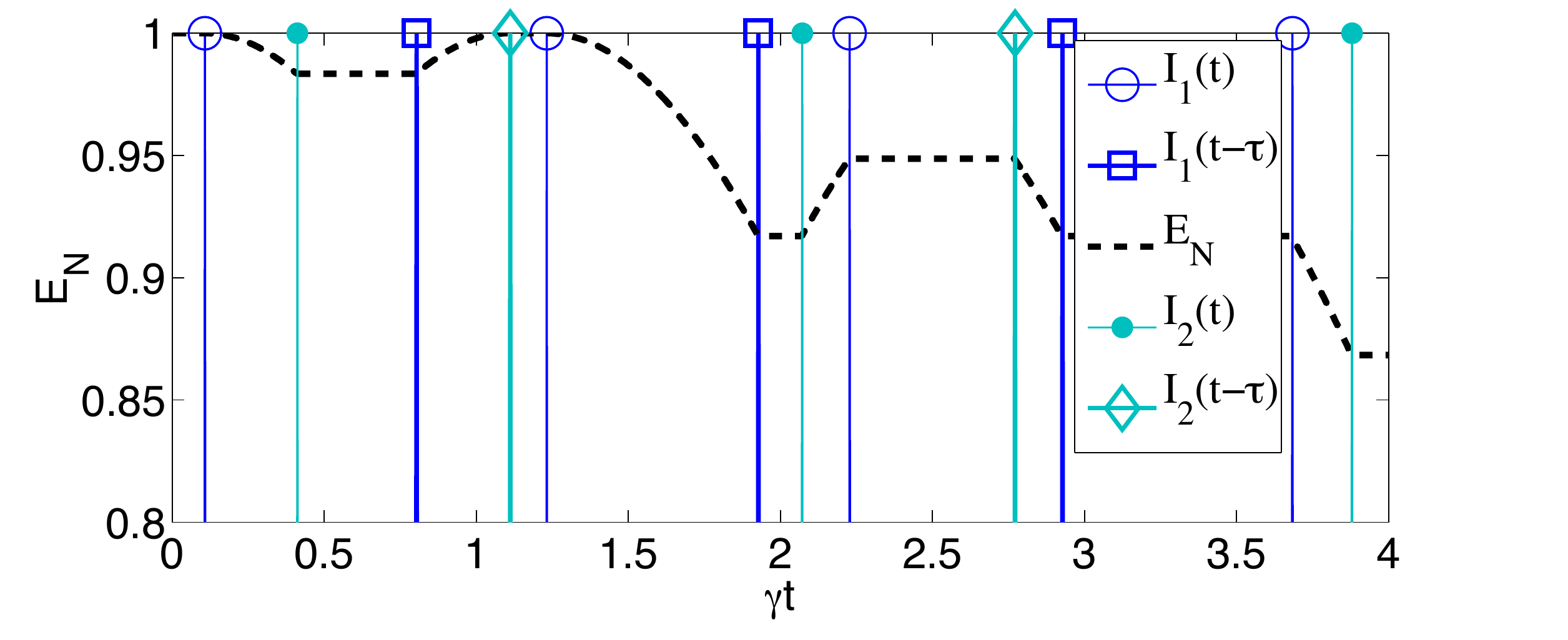}
\caption{Entanglement $E_N$ of a single realization with initial state $|12\rangle+|21\rangle$ and feedback time delay $\tau\approx 0.7/\gamma$. We also show the measurement records in both qutrits and their delayed version indicating the feedback action.}\label{OneTimedelay}
\end{figure}
 This is the best possible situation given the time delay. Entanglement would be completely degraded if a single qutrit emitted two consecutive excitations within the delay time and this would be the worst situation.
In fig. (\ref{OneTimedelay}) we show one particular realization presenting the effects of the finite feedback time delay on the entanglement of a single experimental run. In fig. (\ref{FeedbackTimeDelay}) we show the average (over many trajectories) entanglement as a function of both time and feedback time delay.  
As shown in the figures the greater the time delay the less effective is the recycling process, although a considerable amount of entanglement can still be preserved after long times and an appreciable time delay. It is easy to see that states like $|11\rangle+|22\rangle$ would only be degraded by a time delay.

\begin{figure}[h]
\includegraphics[width=9cm]{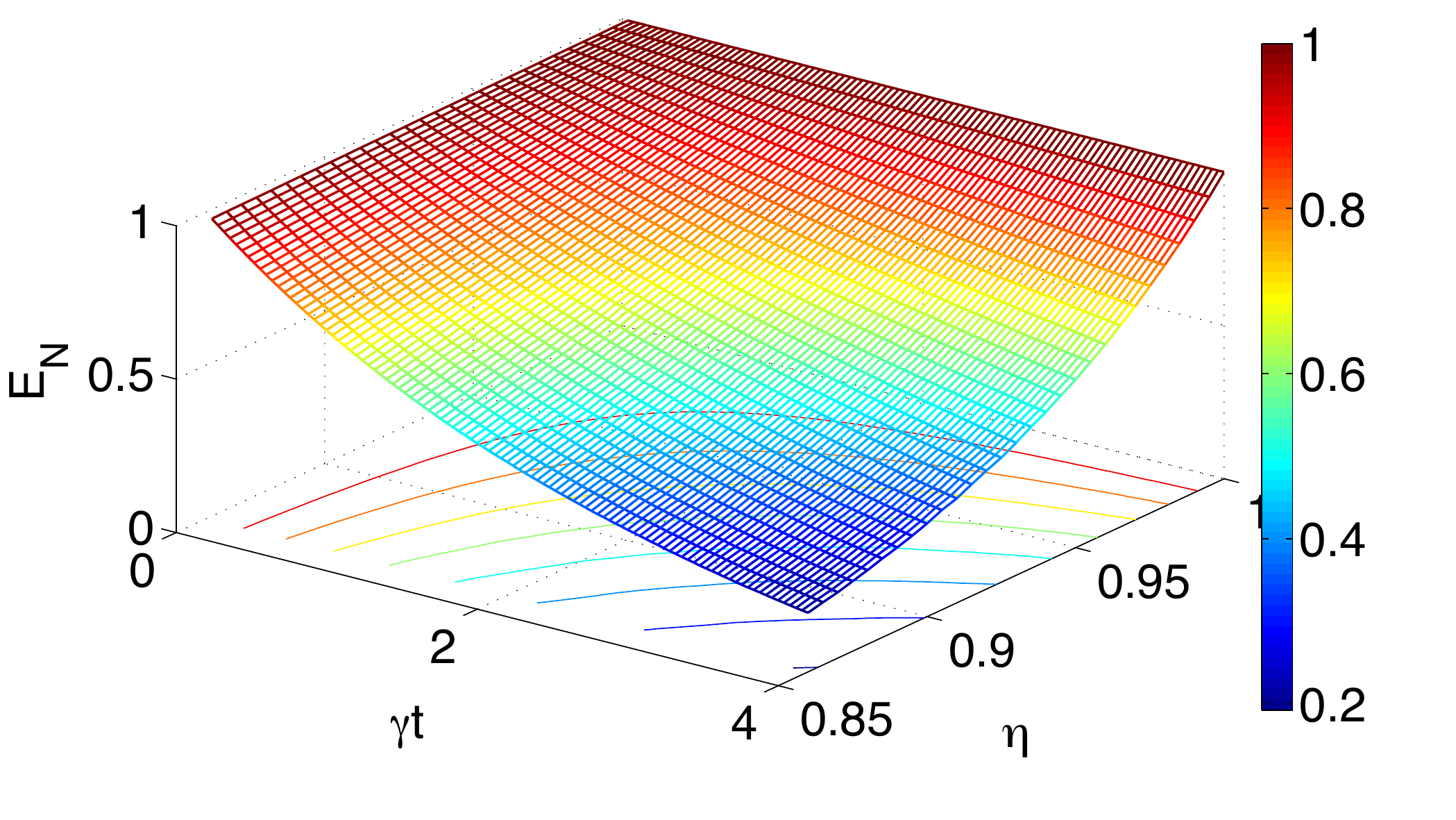}
\caption{Average entanglement (negativity) of the jump feedback as a function of time and measurement efficiency.}\label{MeasIneff}
\end{figure}

\begin{figure}[h]
\includegraphics[width=7cm]{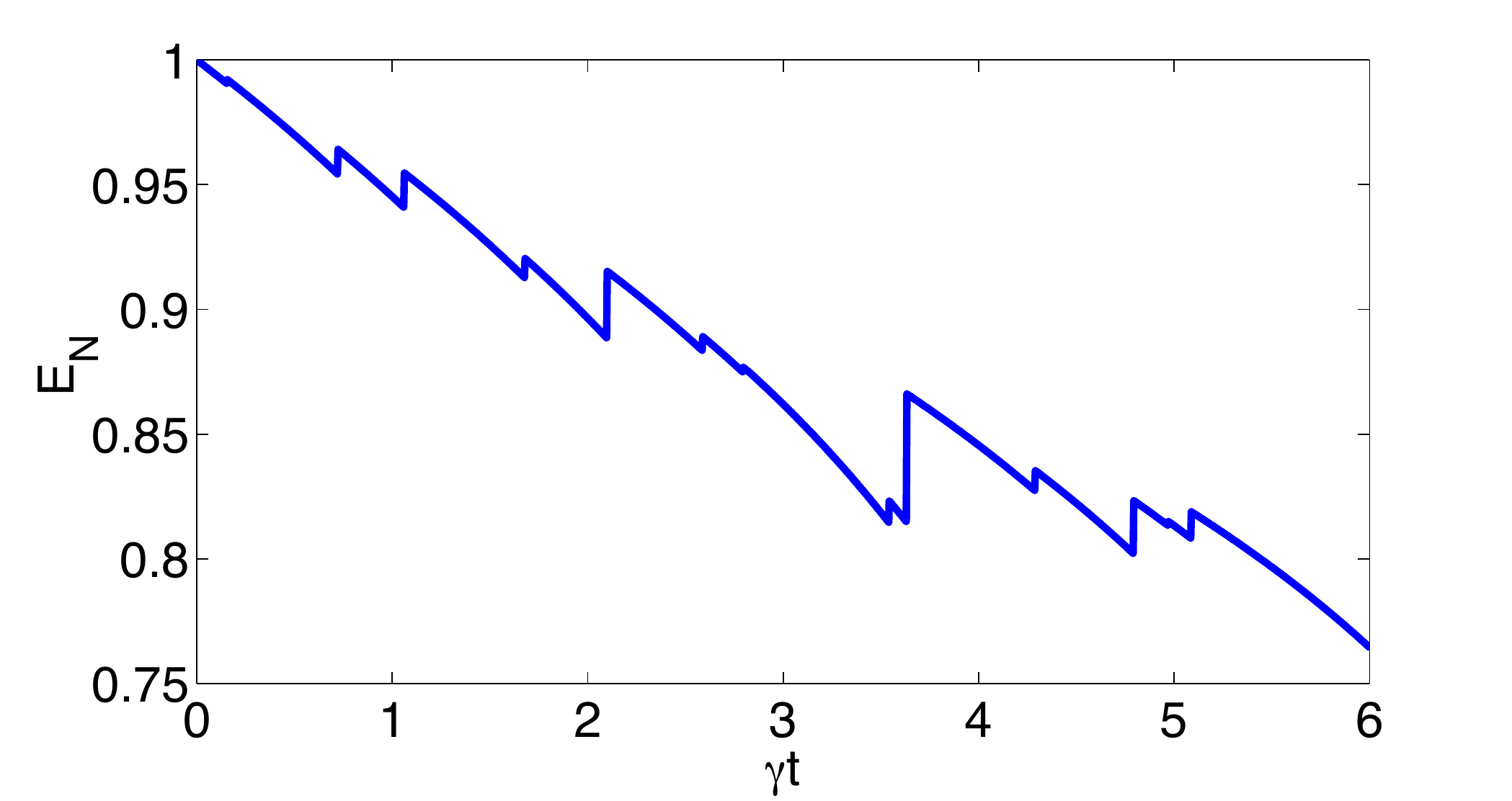}
\caption{Entanglement of a single realization with initial state $|12\rangle+|21\rangle$ and measurement efficiency $\eta\approx 0.98$.}\label{OneMeasIneff}
\end{figure}

The protocol is much more affected by measurement inefficiencies as can be seen in fig. \ref{MeasIneff}. The no jump evolution starts to resemble the unconditioned master equation evolution as the measurement efficiency decreases. It populates the ground state component and decreases the quantum coherence and entanglement in the codespace. As a consequence the recycling process becomes less and less effective as the efficiency decreases. Still an interesting effect which is what makes the protocol effective at small inefficiencies is that when a jump occurs (followed by feedback) the ground state component is eliminated which increases both population and coherence in the codespace leading to an entanglement leap (this effect also played an important role in elucidating a quantum classical transition~\cite{Transition}). Thus in this process a quantum jump always increases entanglement, and this is shown in fig. \ref{OneMeasIneff}.

When the detectors have a non unit efficiency the evolution leads to loss of purity and we generally have a two qutrit mixed state. There is no closed expression for the entanglement (of formation) of such states. Here we choose to plot the negativity~\cite{Nega} as an upper bound to the distillable entanglement. For the sake of comparison we plot the negativity in all the figures.

\begin{figure}[h]
\includegraphics[width=9cm]{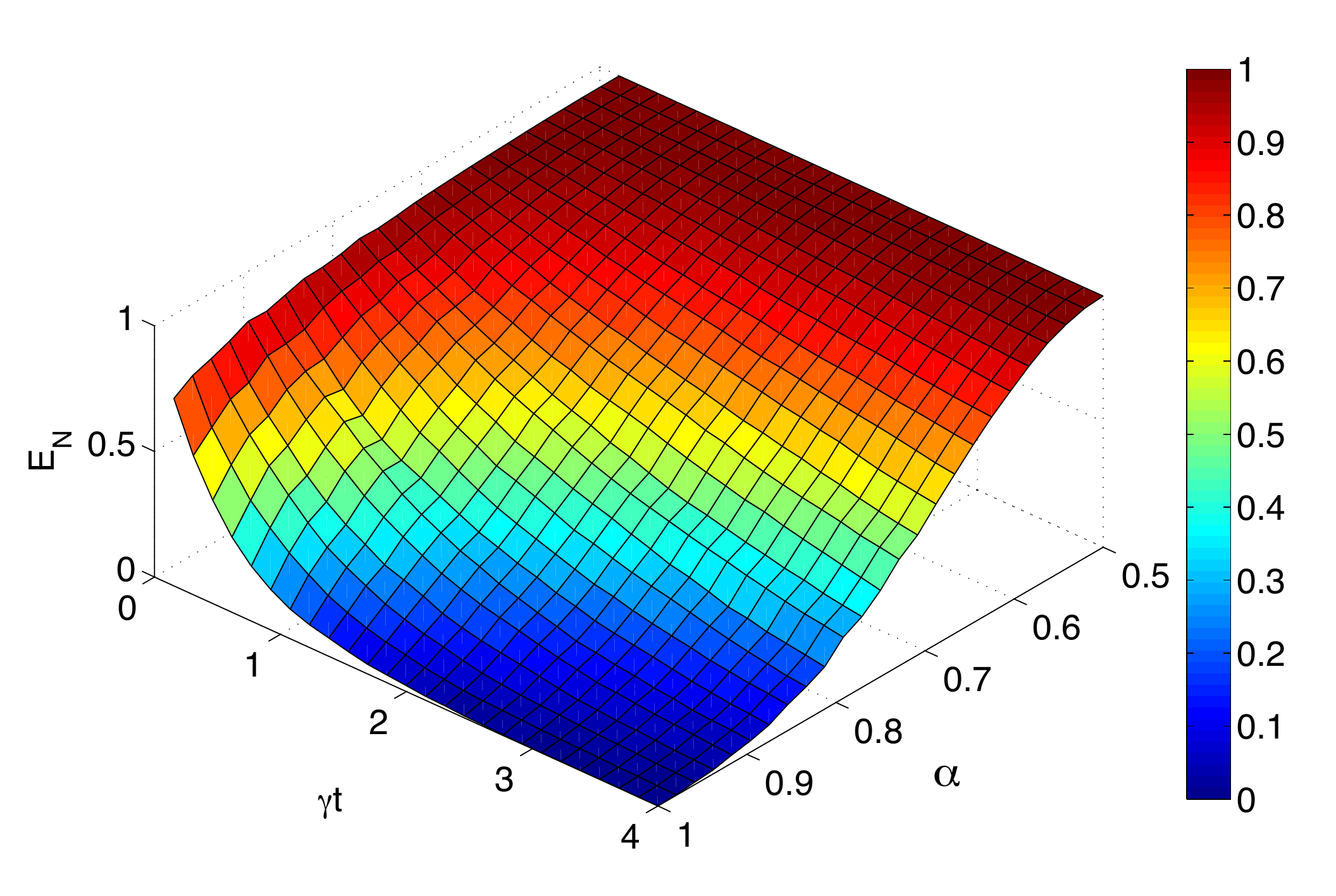}
\caption{Average entanglement (negativity) of the jump feedback as a function of time and distinguishability $\alpha$.}\label{distingui}
\end{figure}
We have stated earlier that the indistinguishability of the decay channels of the qutrit is of extreme importance for the error correction to work. Now, we relax this constraint and look at the protocol effectiveness as the transitions become more and more distinguishable. The single detection operator $\Pi$ is split into two other detectors, $\Pi_2=\sqrt{\gamma}\left[\sqrt{\alpha}|1\rangle\langle2|+\sqrt{1-\alpha}|0\rangle\langle1|\right]$ and $\Pi_1=\sqrt{\gamma}\left[\sqrt{1-\alpha}|1\rangle\langle2|+\sqrt{\alpha}|0\rangle\langle1|\right]$, each one corresponding to one of the transitions. If $\alpha=1/2$ the transitions are indistinguishable and if $\alpha=1$ (or zero) the transitions are perfectly distinguishable. Note that by changing the distinguishability we also change the overall dynamics (not just the measurement process), such that for $\alpha\neq1/2$ we have $\mathcal{L}[\Pi]\neq\mathcal{L}[\Pi_1]+\mathcal{L}[\Pi_2]$, however for $\alpha=1/2$ we have $\mathcal{L}[\Pi]=\mathcal{L}[\Pi_1]+\mathcal{L}[\Pi_2]$. After both detectors are triggered the same feedback action is applied. As shown in figure (\ref{distingui}) the protocol is still highly effective if the transitions are slightly distinguishable. However, as the transitions become more distinguishable the errors are dominant and the system loses entanglement quickly. 

\begin{figure}[h]
\includegraphics[width=9cm]{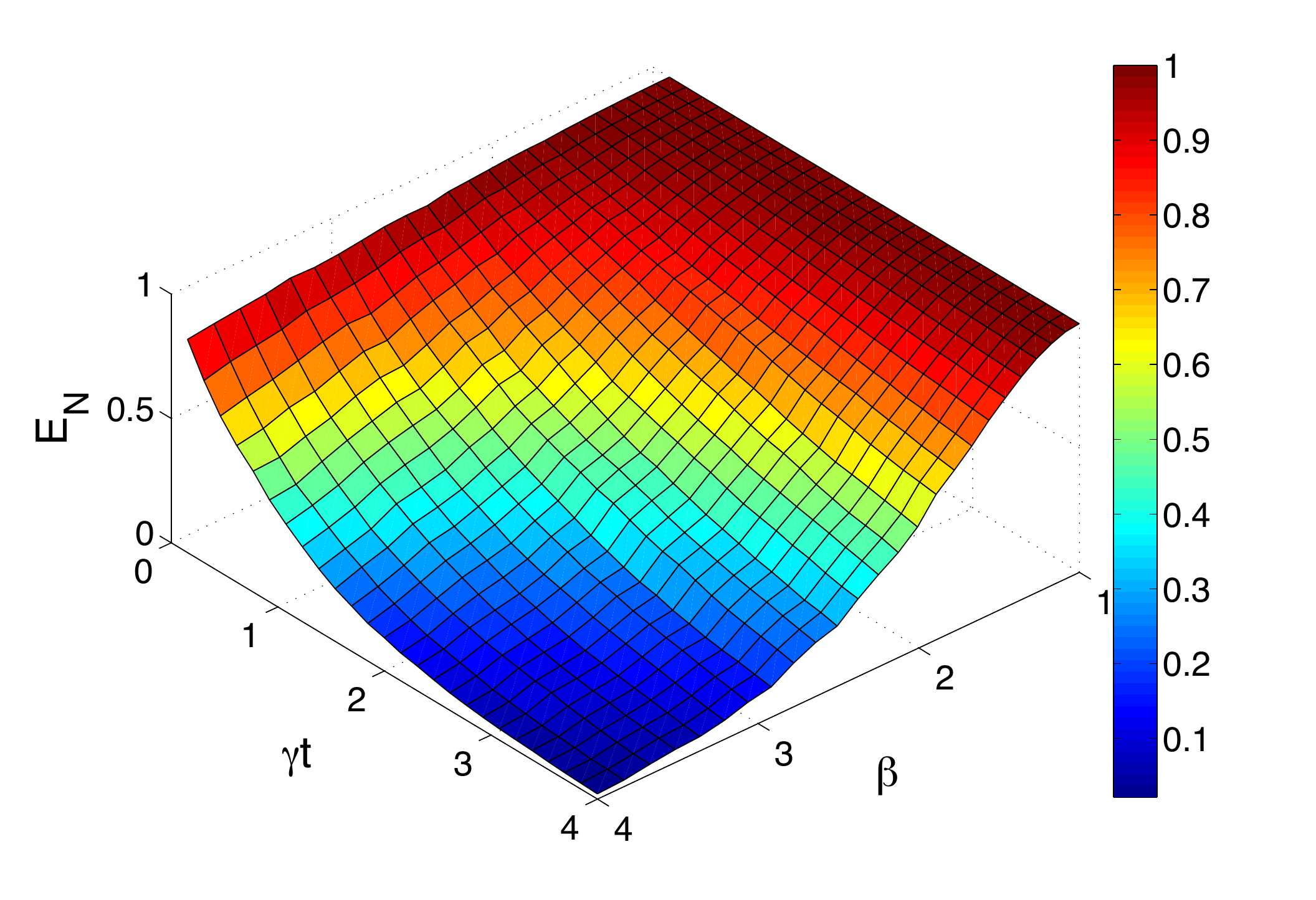}
\caption{Average entanglement (negativity) of the jump feedback as a function of time and the transition rate $\beta$.}\label{taxas}
\end{figure}
Even though the transitions may be indistinguishable the transition rates may differ, which makes one of them more probable than the other. This is the case in harmonic oscillators, for instance. Generically the jump operator would be given by $\Pi=\sqrt{\gamma}[\sqrt{\beta}|1\rangle\langle2|+|0\rangle\langle1|]$, yielding a probability for the transition $|2\rangle\rightarrow|1\rangle$ proportional to $\beta$. When the rates are unbalanced the no jump (or no detection) evolution disturbs the chosen codespace, which means that our strategy could be combined with a "bang-bang" error correction protocol~\cite{BangHybrid, bang} to protect the logical qubits (although we do not apply the "bang-bang" control). Furthermore, the jump evolution also disrupts the encoded information in addition to taking it outside the codespace. As the upper transition becomes more and more probable the  protocol becomes less effective. However, for unbalanced transition rates corresponding to harmonic oscillators ($\beta=2$) the protocol is still able to protect a considerable amount of entanglement, as shown in fig. (\ref{taxas}).

\begin{figure}[h]
\includegraphics[width=8.5cm]{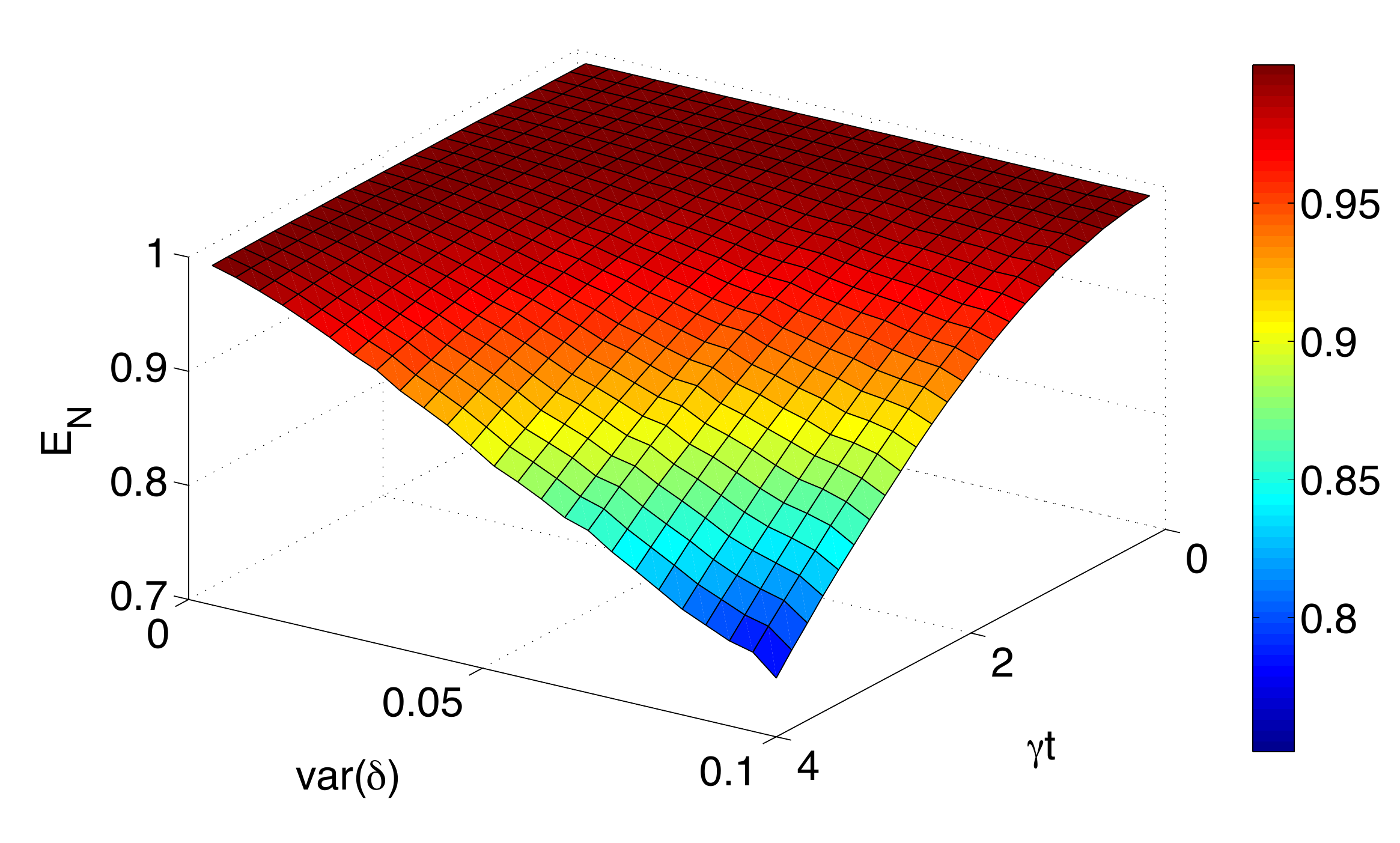}
\caption{Average negativity as a function of time and fluctuations in the Hamiltonian strength. The fluctuations are quantified by the variance $var(\delta)$ of the Gaussian distribution centered at the ideal Hamiltonian strength $\lambda$}\label{FeeDisord}
\end{figure}

Another possible imperfection in the feedback mechanism is an imprecision in the implementation of the recycling operation, such as
a fluctuating Hamiltonian strength as in a disorder effect. For instance the Hamiltonian could be given by the ideal one with a strength which assumes Gaussian distributed values around the ideal value $\lambda$ each time the feedback is triggered. The disordered feedback Hamiltonian would be given by $(1+\frac{\delta}{\lambda})F$, with $\delta$ a Gaussian distributed random variable with zero mean. This disorder effect amounts to a non perfect recycling operation that is subjected to random imperfections and it is given by a one parameter family of random unitary matrices: the ideal operation times a one parameter family $Re^{-iF\frac{\delta}{\lambda}}$. This disorder recycling process often does not restore the state back to the codespace since the disordered matrices may differ considerably (for an appreciable variance of the disorder distribution $var(\delta)$) from the ideal, thus leading to a degradation of global resources during periods in which the system ceases emission. We show the average negativity in fig. (\ref{FeeDisord}). It can be seen that the protocol tolerates small fluctuations in the Hamiltonian and we believe this disorder effect to be not as significant as measurement inefficiency and time delay, although for higher disorder the effect would be far more damaging than a time delay since the average entanglement decreases approximately linearly with the time delay and presents a decreasing curvature with the disorder. 

 \section{Distinguishable decay channels and qutrit disentanglement}
 
So far, we have shown that by including a third level in the usual qubit protocols we may actually prevent losses with, relatively robust, local strategies, which was not possible by working strictly with qubits. The idea is based on cascade decay and the indistinguishability of the excitations emitted by different decay channels. However, one could ask what happens when those channels are completely distinguishable and therefore entanglement cannot be protected by the local recycling protocol. In this session we address this issue by looking at the dynamics of entanglement of $3 \otimes 3$ systems for different internal structures.  Note that in our analysis, each system will always be subjected to two independent decay channels. As we have seen, for a decaying qubit the only possible structure for the decaying channel is a transition from the excited state to the ground state and the only possible behaviors of the disentanglement dynamics are an asymptotic or a sudden death of entanglement~\cite{ESuddenD}. We show in this session that for qutrits, on the other hand, the picture is richer and may present asymptotic entanglement even under the action of local independent reservoirs. 

\begin{figure}[h]
\includegraphics[height=2.3cm]{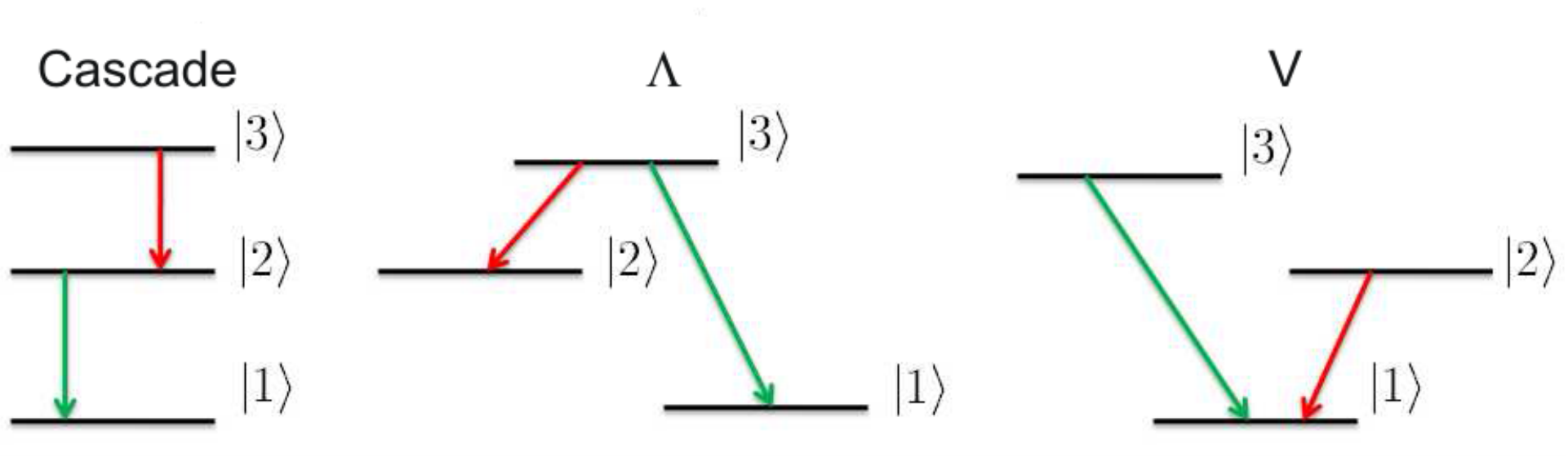}
\caption{Qutrit decaying structures with one forbidden or low probability transition.}
\label{structures}
\end{figure}

We study three different distinguishable decay channel configurations for the qutrits:\\
1) The cascade or E structure with jump operators 
\begin{eqnarray}
\nonumber
E_1=\sqrt{\gamma_{1}}|0\rangle\langle 1|\\ 
E_{2}=\sqrt{\gamma_{2}}|1\rangle\langle 2|
\end{eqnarray}
2) the V structure with
\begin{eqnarray}
\nonumber
V_{1}=\sqrt{\gamma_{1}}|0\rangle\langle 1|\\
V_{2}=\sqrt{\gamma_{2}}|0\rangle\langle 2|
\end{eqnarray}
3) the $\Lambda$ configuration with
\begin{eqnarray}
\nonumber
\Lambda_{1}=\sqrt{\gamma_{1}}|0\rangle\langle 2|\\
\Lambda_{2}=\sqrt{\gamma_{2}}|1\rangle\langle 2|
\end{eqnarray}
as shown in Fig.~\ref{structures}.

As we show bellow, different configurations will present different dynamics. Before we proceed, please note that, throughout the analysis we will always consider independent reservoirs of the same strength, $\gamma_{1}=\gamma_{2}$, with no loss of generality (all the important ingredients already show up in this simple case) and reservoir detectors that are able to distinguish photons emitted by each internal transition hence avoiding the addition of any externally induced entanglement in the system. Finally, we also focus on initial states of the form $|\psi\rangle=a|00\rangle+b|11\rangle+c|22\rangle$ (where $a$, $b$ and $c$ are real numbers and $a\neq b\neq c\neq 0$ for simplicity), which, once again, with no loss of generality, already present all the interesting features in qutrit-qutrit disentanglement.

\begin{figure}[h]
\includegraphics[height=7.0cm]{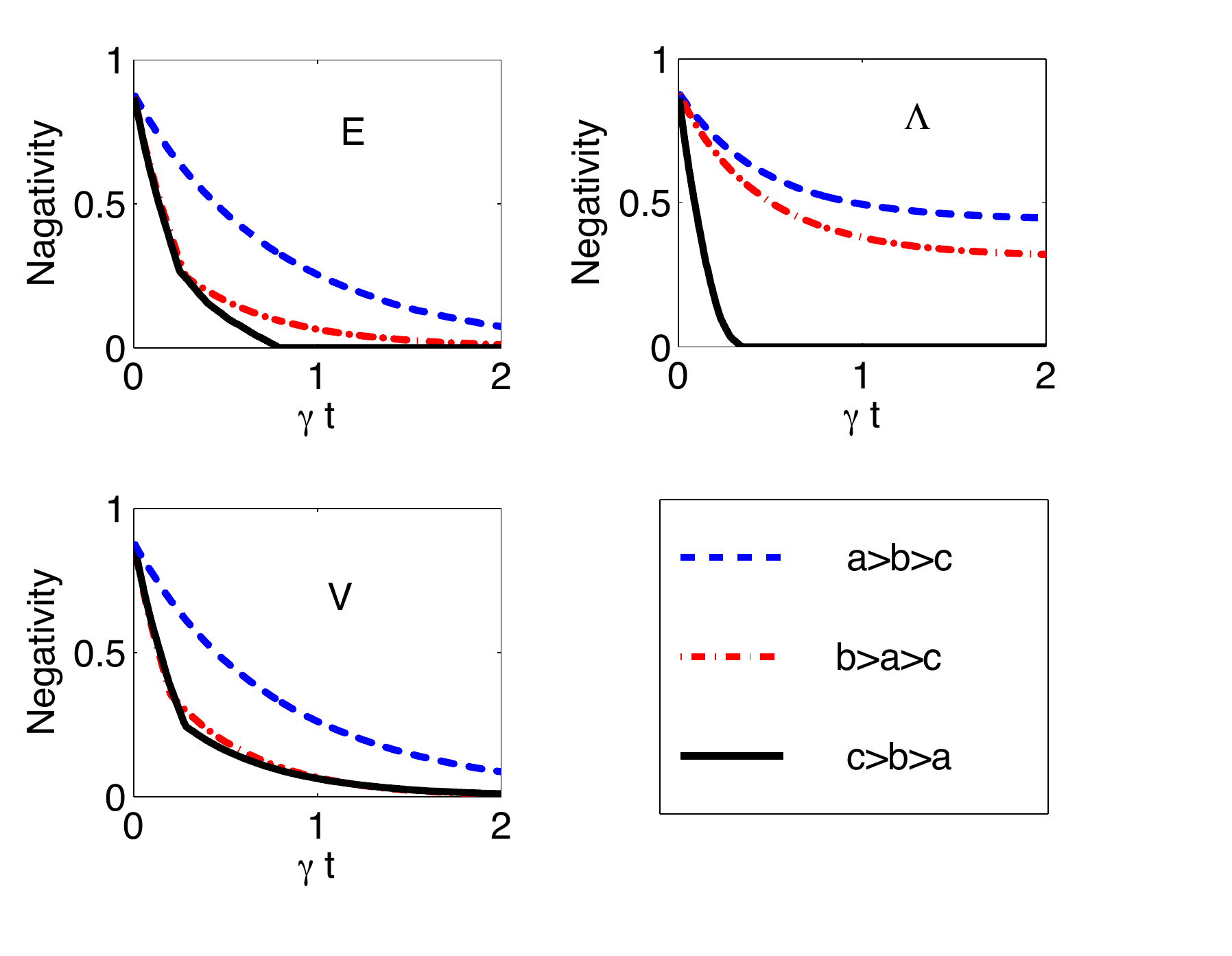}
\caption{Examples summarizing the disentanglement dynamics for the configurations $E,\Lambda$, and $V$ for different initial conditions $(a,b,c)$. }
\label{qutritrho}
\end{figure}

Previous works have introduced the possible elements of the dynamics of disentanglement in $3 \otimes 3$ systems under dissipation such as precursors of sudden death, sudden death of distillability, and asymptotic decay~\cite{Abrupt}. All these elements are present here but not necessarily for all the configurations. Again, as in previous cases, the presence of those elements also depend on the initial state. We summarize bellow the relations between the decaying structure, the initial state, and the resulting general aspects of the disentanglement dynamics for each internal structure (and show some examples in Fig. \ref{qutritrho}):
\begin{enumerate}
\item in cascade - cascade structure: (a) aymptotic decay for $a>b>c$, (b) one sudden change followed by asymptotic decay for $b>a>c$ or $a>c>b$, (c) two sudden changes and asymptotic decay for $b>c>a$, and (d) two sudden changes and sudden death for $c>b>a$;
\item in V - V structure: (a) aymptotic decay for $a>b$ and $a>c$, (b) one sudden change and asymptotic decay for $a>b$ and $a<c$, and (c) two sudden changes and asymptotic decay for $a<b$ and $a<c$;
\item in $\Lambda$ - $\Lambda$: (a) asymptotic decay with asymptotic entanglement for $2a>c$ and $2b>c$, (b) one sudden change and asymptotic decay with asymptotic entanglement for $2a>c$ and $2b<c$ with $c^{2}<4ab$, (c) two  sudden change and asymptotic decay for $2a>c$ and $2b<c$ with $c^{2}>4ab$, and (d) two sudden change and sudden death for $2a<c$ and $2b<c$.
\end{enumerate}

\begin{figure}[h]
\includegraphics[height=7.0cm]{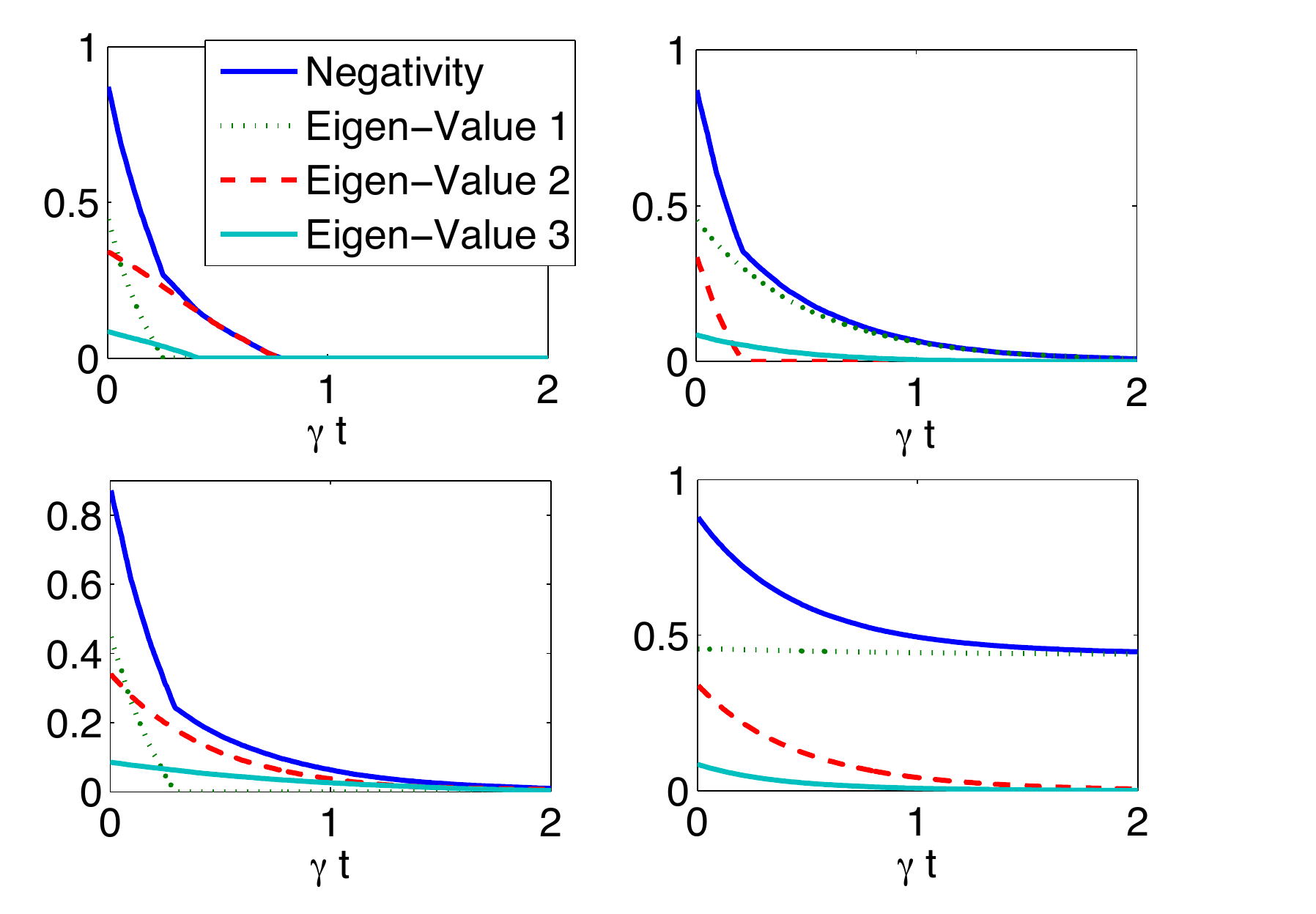}
\caption{Examples of disentanglement dynamics showing the negative eigenvalues of the partial transposition of the density matrix, such that a sudden change occurs every time a negative eigenvalue vanishes (or becomes positive). (a)two abrupt changes and sudden death in Cascade structure, and (b) two abrupt changes and asymptotic decay in V structure ($a=0.179$, $b=0,2386$ and $c=0,9545$); (c) one abrupt change in Cascade structure and (d) no abrupt changes and asymptotic non null entanglement in $\Lambda$ structure ($a=0.2386$, $b=0,9545$ and $c=0,1790$). }
\label{qutritrho1}
\end{figure}

We also show the negative eigenvalues of the partial transposition of the density matrix for some examples, recalling that a sudden change occurs every time one of the negative eigenvalues becomes non negative (fig. \ref{qutritrho1}). 

In all the cases we study, entanglement is measured by negativity which for $3 \otimes 3$ systems estimates the amount of free entanglement available in the system~\cite{Bound}. The first striking feature present in Fig. \ref{qutritrho} is that in the $\Lambda - \Lambda$ configuration, entanglement may not decay to zero but to a non-null value, depending on the initial state (if $c^{2}<4ab$). This interesting property can be explained by the fact that there is indeed an entangled dark state in this configuration, composed of the superposition of the product of the lower states in each part, which is not affected by the dissipative dynamics. 

Another interesting feature is that in the V structure only when $b$ or $c=0$ the system may present entanglement sudden death. In this case, we have effectively a two-qubit system, given that one of the upper levels of each qutrit is initialy not populated and, at least for zero temperature reservoirs, will remain empty for all times. If, on the other hand, both $b$ and $c$ are different from zero, then there will always be a residual amount of entanglement in the $\{|11\rangle,|22\rangle\}$ subspace intil all the coherence of the initial superposition dies, which only happens asymptotically in time, hence, no entanglement sudden death. Naturally, if the temperature of the reservoir is different from zero then the system is incoherently pumped and entanglement  dies in finite time. 

\begin{figure}[h]
\includegraphics[height=5cm]{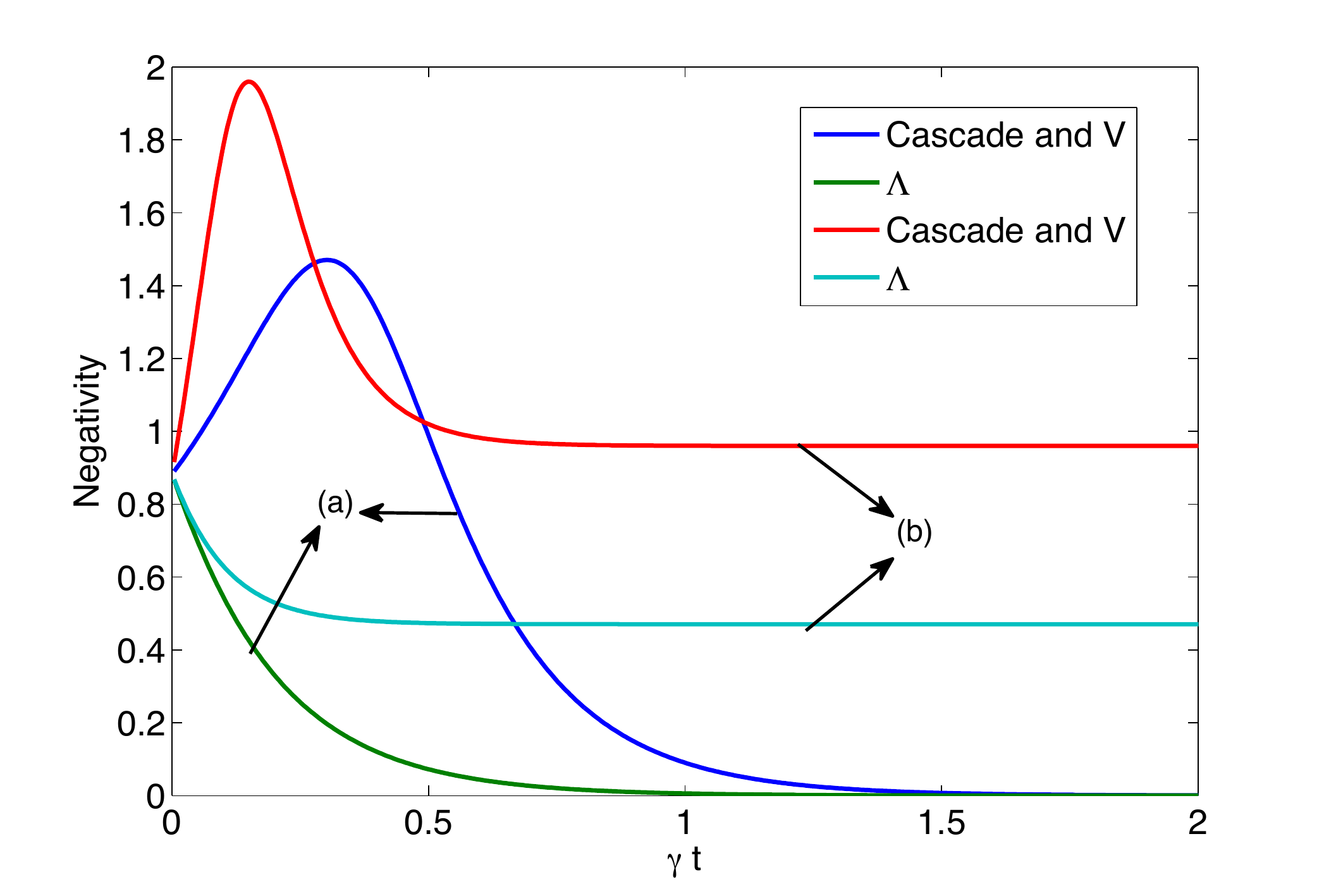}
\caption{ The figures shows the two possible entanglement evolution in no jump trajectories in (a) V and Cascade configuration and (b) Lambda Configuration (a = 0:1790, b =0; 2386  and c = 0; 9545 in trajectories whose entanglement increases;
and a = 0:9545, b = 0; 2386 and c = 0; 1790 in trajectories whose entanglement decays). }
\label{njump}
\end{figure}

Much in the same way as in the previous sessions, we can also analyze the effects of continuous monitoring of the reservoirs in the dynamics of the entanglement of the system, now as a function of the configuration of the levels. However, this time we only focus on no jump trajectories since all transition-emitted photons are distinguishable and a single decay completely identifies the quantum state of the decaying qutrit and definitely kills entanglement. When the reservoirs are monitored and no decay is detected, the general evolution of the system will be given by
\begin{equation}
|\psi(t)\rangle=\frac{|\tilde{\psi}(t)\rangle}{\sqrt{\langle\tilde{\psi}(t)|\tilde{\psi}(t)\rangle}},
\end{equation} 
with $|\tilde{\psi}(t)\rangle=e^{H_{eff}t}|\psi(0)\rangle$,
where $|\psi(0)\rangle$ is the initial state and $H_{eff}$ is the effective Hamiltonian (of no detection) given by $H_{eff}=-\frac{1}{2}\sum_n \Pi^{\dagger}_n\Pi_n$, with $\Pi_n$ being the jump operators. 
Incidently, $E$ and $V$ configuration present the same effective Hamiltonian, consequently, the same dynamics. For the initial states under consideration, the non-normalized state and the negativity are respectively given by $
|\tilde{\psi}(t)\rangle=a|00\rangle+b e^{-2\gamma t}|11\rangle+ce^{-2\gamma t}|22\rangle$ and
\begin{equation}E_N=\frac{2}{a^2+(b^{2}+c^{2})e^{-4\gamma t}}(abe^{-2\gamma t}+ace^{-2\gamma t}+bce^{-4\gamma t}).
\end{equation}
In such no-jump trajectories there is also an effect reminiscent of the optimum singlet state conversion for qubits~\cite{Nos}. Starting with a non maximally entangled state there is the possibility of increasing entanglement before it reaches the asymptotic value. But now the system may attain higher values of entanglement since it is a qutrit-qutrit and not a qubit-qubit system (fig.~(\ref{njump})).

The $\Lambda$ structure on the oder hand presents a different effective Hamiltonian.
In this case not only it is possible to increase the entanglement by measurement, but there is also the possibility of reaching an asymptotic entangled state. The no-jump evolution is given by
$
|\tilde{\psi}(t)\rangle=a|00\rangle+b|11\rangle+ce^{-4\gamma t}|22\rangle$ and the negativity is then
\begin{equation}
E_N=2\frac{ab+(a+b)ce^{-4\gamma t}}{1+c^{2}(e^{-8\gamma t}-1)}.
\end{equation}
In the asymptotic limit the negativity is $N(t\rightarrow\infty)=\frac{2ab}{a^{2}+b^{2}}$.
Note that, this behavior does not depend on decay rates. The asymptotic entanglement in no-jump and unconditional dynamics occurs because the $\Lambda$ structure presents a $2\otimes2$ decoherence free subspace ($|00\rangle$, $|01\rangle$, $|10\rangle$ and $|11\rangle$), as a two-qubit system (figure~(\ref{njump})).
The simple monitoring scheme is then able to prevent the abrupt changes, including sudden death of distillability, and in fact, replace them with a transient increase of entanglement followed by an asymptotic preservation of entanglement through only local operations. 

\section{Conclusions}

We conclude by stressing that the analysis done in this work has identified different limitations as well as ex- tended the possibilities of using qutrits to encode qubits in dissipative single copy quantum protocols. In particular, we have shown that in the photon detection limit the
encoding is particularly sensitive to the efficiency of the measurement process, whereas quite robust against time delay in the feedback mechanism and very robust against fluctuations in the feedback interaction. We have also studied the limit corresponding to homodyne detection and have shown that the encoding in qutrits allows for a certain degree of protection of the initial entanglement which is not possible at all in this regime if the qubits are encoded in two-level systems. Finally, we have analyzed the scenario in which the decay channels are orthogonal, hence single clicks already kill the entanglement of the system. We have studied different qutrit configurations and we have shown that in some of them the no jump trajectory may still preserve entanglement even asymptotically in time due to the presence of decoherence free subspaces in the system.

\acknowledgements
We would like to acknowledge D. Cavalcanti for useful discussions and the support from the Brazilian agencies CNPq and Fapemig. This work is part of the Brazilian National Institute of Science and Technology on Quantum Information (INCT-IQ). E. M. acknowledges Alexia Auff\`eves and the N\'eel institute in Grenoble for their hospitality.

\end{document}